\documentclass[preprint,showpacs,preprintnumbers,amsmath,amssymb]{revtex4}

\usepackage{graphicx}
\usepackage{dcolumn}
\usepackage{bm}
\usepackage[none]{hyphenat}

\begin{document}

\preprint{APS/123-QED}

\title{Measurements of proton induced reaction cross sections on $^{120}$Te  \\
for the astrophysical $p$-process \\ }

\author{R. T. G\"{u}ray\footnote{Corresponding Author : tguray@kocaeli.edu.tr }}%

\author{N. \"{O}zkan}
\author{C. Yal\c{c}{\i}n}
\affiliation{Kocaeli University, Department of Physics, Umuttepe
41380, Kocaeli, Turkey
}%
\author{A. Palumbo}
\author{R. deBoer}
\author{J. G\"{o}rres}
\author{P. J. LeBlanc}
\author{S. O'Brien}
\author{E. Strandberg}
\author{W. P. Tan}
\author{M. Wiescher}
\affiliation{University of Notre Dame, Department of Physics, Notre
Dame, Indiana 46556, USA
}%
\author{Zs. F\"{u}l\"{o}p}
\author{E. Somorjai}
\affiliation{ATOMKI, H-4001 Debrecen, POB.51., Hungary
}%
\author{H. Y. Lee}
\author{J. P. Greene}
\affiliation{Argonne National Laboratory, Illinois 60439, USA
}%

\date{\today}

\begin{abstract}

The total cross sections for the $^{120}$Te(p,$\gamma$)$^{121}$I
and $^{120}$Te(p,n)$^{120}$I reactions have been measured by the
activation method in the effective center-of-mass energies 2.47
MeV $\leq$ $E_{\rm c.m.}^{eff}$ $\leq$ 7.93 MeV and $\mbox{6.44
MeV $\leq$ $E_{\rm c.m.}^{eff}$ $\leq$ 7.93 MeV,}$ respectively.
The targets were prepared by evaporation of $\mbox{99.4\%}$
isotopically enriched $^{120}$Te on Aluminum and Carbon backing
foils, and bombarded with proton beams provided by the FN tandem
accelerator at the University of Notre Dame. The cross sections
and $S$ factors were deduced from the observed $\gamma$ ray activity,
which was detected off-line by two Clover HPGe detectors mounted
in close geometry. The results are presented and compared with the
predictions of statistical model calculations using the codes
NON-SMOKER and TALYS.

\end{abstract}

\pacs{{25.40.Lw}, {26.30.-k}, {27.60.+j}
}
\maketitle

\section{\label{sec:level1}Introduction}

The elements heavier than iron \mbox{($Z >$ 26)} are mainly synthesized
by three mechanisms: the $s$-process, $r$-process, and
$p$-process, the latter being the least known among them. The
$p$-process, responsible for the production of 35 proton-rich
stable isotopes, can proceed via a combination of
photodisintegration reactions  -\mbox{($\gamma$,$n$)},
\mbox{($\gamma$,$p$)} and \mbox{($\gamma,\alpha$)}- on existing
heavy $s$- and $r$-seeds in the temperature range of \mbox{2 - 3 x
10$^9$ K}. These high temperatures can be achieved in explosive
environments, such as the O/Ne layers of $\mbox{Type-II}$
supernovae \cite{{Woosley}, {Rayet90}}. Initially, the nuclides
are driven by a sequence of \mbox{($\gamma$,$n$)} reactions to the
proton-rich side of the valley of stability, whereby the binding energies of neutrons
gradually increase along the isotopic path. When the
\mbox{($\gamma$,$p$)} and/or \mbox{($\gamma,\alpha$)} reaction
rates become significant compared to those of the ($\gamma$,$n$),
the reaction path branches towards lower $Z$ nuclei. While the
photodisintegration reactions govern the overall reaction flow,
complementary processes such as $\beta^{+}$ decays, electron
captures, and \mbox{($n$,$\gamma$)} reactions may play an important role as
well. Recent p-process simulations demonstrate that
\mbox{($\gamma$,$\alpha$)} reactions determine the overall
reaction flow in between closed shells and affect the medium and
heavy $p$-nuclei abundances. On the other hand,
\mbox{($\gamma$,$p$)} reactions provide important links for
feeding $p$-process nuclei \cite{Rapp06}. In particular, $^{120}$Te
is populated by a sequence of two photodisintegration
reactions $^{122}$Xe($\gamma,p$)$^{121}$I($\gamma,p$)$^{120}$Te.
Simulations indicate that $^{120}$Te is underproduced in
comparison to $p$-process abundance observations \cite{Rapp06} which
supports the results of earlier calculations \cite{{Rayet95},
{Arnould}}.

These $p$-process simulations rely on complex network calculations
including more than 20000 reactions on about 2000 mostly unstable
nuclei. Most of the reaction rates involved in these simulations
are based on statistical model or Hauser-Feshbach predictions. The
overall reliability of Hauser-Feshbach predictions in $p$-process
simulations has been discussed \cite{Arnould}. Variations can lead
to substantial changes in $p$-process abundance predictions. There
is considerable effort to experimentally test the reliability of
$p$-process predictions on selected cases \cite{{Som98}, {Ozkan07}, {Gyu06}, {Yalcin09}, {Kiss07}, {Gyu03}, {Gyu01}, {Galan03}, {Tsagari04}, {Laird87}, {Har01}, {Sauter97}, {Chloupek99}, {Bork98}, {Ozkan01}, {Ozkan02}, {Gyu07}, {Famiano08}, {Kiss08}, {Spy08}}. Since the photodisintegration process $^{121}$I($\gamma,p$) directly feeds
$^{120}$Te, this reaction provides an excellent case for testing
the reliability of the Hauser-Feshbach prediction near the closed
proton shell Z=50; $^{121}$I has a high level density near the
proton threshold and the reaction should be eligible for the
statistical model prediction \cite{Raus97}.

A direct measurement of the \mbox{$\gamma$-induced}
photodisintegration $^{121}$I($\gamma,p$)$^{120}$Te is difficult
since it requires photodisintegration of a short-lived radioactive
$^{121}$I isotopes. Direct \mbox{($\gamma$,$p$)} and
\mbox{($\gamma$,$\alpha$)} photodisintegration measurements have
been demonstrated successfully \cite{{Raus06}, {Utsunomiya06},
{Mohr07}} and $p$-process reactions with Coulomb dissociation
techniques will be pursued at future radioactive beam facilities.

Presently the applicability of the statistical model approach is
limited to testing the inverse radiative capture reaction process.
A measurement of the radiative capture
$^{120}$Te($p,\gamma$)$^{121}$I and the nuclear
$^{120}$Te($p,n$)$^{120}$I reactions does not provide complete
information about the reverse photodisintegration process but it
is suitable for testing the reliability of the Hauser-Feshbach
predictions \cite{Raus00} for these reaction channels. On the
basis of the resulting statistical model parameters, the cross
section predictions for the photodisintegration process can be
directly deduced in the framework of the Hauser-Feshbach model
\cite{Raus04}.

The last decade has seen an increased interest in measuring
the proton capture cross sections of $p$-nuclei \cite{{Kiss07}, {Gyu03}, {Gyu01}, {Galan03}, {Tsagari04}, {Laird87}, {Har01}, {Sauter97}, {Chloupek99}, {Bork98}, {Ozkan01}, {Ozkan02}, {Gyu07}, {Famiano08}, {Kiss08}, {Spy08}}. The bulk of these
\mbox{($p$,$\gamma$)} measurements have been done in the lower
mass region ($A$ $<$ 100).  The cross sections determined in these
\mbox{($p$,$\gamma$)} measurements generally agree with the
statistical model predictions within a factor of two. In contrast,
very few \mbox{($p,n$)} measurements have been performed, again
mainly on nuclei in the lower mass range of the $p$-process such as
$^{76}$Ge \cite{Kiss07}, $^{82}$Se \cite{Gyu03}, and $^{85}$Rb
\cite{Kiss08}. Measurements on $^{120}$Te expand the range of
$p$-process studies and address a case where the feeding process
$^{121}$I($\gamma,p$)$^{120}$Te of a $p$-nucleus can be studied.

The charged-particle reaction cross sections for
$^{120}$Te($p,\gamma$)$^{121}$I and $^{120}$Te($p,n$)$^{120}$I can
be measured via the activation technique since in both cases the
products are radioactive and have appropriate
\mbox{$\beta$-decay} half-lives.  In the case of
\mbox{$^{120}$Te($p,n$)$^{120}$I}, the product $^{120}$I has ground
($^{120g}$I) and isomeric states ($^{120m}$I), and their partial
cross sections can be determined separately because of the
different decay pattern of the two states. The decay parameters
used for this analysis are summarized in Table I. The details of
the experiment are given in Sec. II.

The \mbox{$^{120}$Te($p$,$\gamma$)$^{121}$I} and
\mbox{$^{120}$Te($p,n$)$^{120}$I} activation measurements have
been performed up to \mbox{7.93 MeV} as a test of the statistical
model predictions over a broader energy region. The experimental
cross sections have been compared with the predictions of
Hauser-Feshbach statistical model calculations using the codes
standard NON-SMOKER and TALYS with various combinations of the
nuclear inputs. This analysis is discussed in Sec. III. A summary
and conclusions are provided in Sec. IV.

\section{Measurements}

\subsection{Target properties}

Targets were prepared at Argonne National Laboratory and at the
University of Notre Dame. \mbox{99.4\%} enriched $^{120}$Te oxide
was evaporated onto \mbox{20 $\mu$g cm$^{-2}$} C backing
\cite{Greene08} and \mbox{1.5 mg cm$^{-2}$} Al backing, and two
targets were produced with thicknesses of \mbox{128 $\mu$g
cm$^{-2}$} and \mbox{456 $\mu$g cm$^{-2}$}, respectively. Target
frames were made of Ta, with \mbox{1 cm} diameter holes. Target
thicknesses were checked by Rutherford backscattering (RBS) and
verified to within \mbox{9\%} uncertainty.

Backscattered protons were measured in order to monitor the target
performance during each irradiation. For this purpose, a Si surface
barrier detector with a reduced entrance aperture of \mbox{0.5 mm}
diameter was mounted at an angle of 135$^\circ$ with respect to
the beam direction as shown in \mbox{Fig. 1}.

\begin{table*}
\caption{ Decay parameters of the $^{120}$Te\,+\,$p$ reaction
products \cite{nudat} and measured photo-peak efficiencies of the
$\gamma$ transitions, used for the analysis.}
\setlength{\extrarowheight}{0.1cm}
\begin{ruledtabular}
\begin{tabular}{cccccc}
\parbox[t]{2.0cm}{\centering{Reaction}} &
\parbox[t]{1.5cm}{\centering{Product}} &
\parbox[t]{2.0cm}{\centering{Half-life }} &
\parbox[t]{2.0cm}{\centering{$\gamma$ Energy (keV)}} &
\parbox[t]{2.0cm}{\centering{$\gamma$ Intensity (\%)}}&
\parbox[t]{3.0cm}{\centering{  Detection efficiency (\%)}} \\
\hline
$^{120}$Te($p$,$\gamma)$& $^{121}$I & (2.12 $\pm$ 0.01) h & 532.08 &6.1 $\pm$ 0.3& 12.6 $\pm$ 0.3\\
$^{120}$Te($p,n$)& $^{120g}$I & (81.6 $\pm$ 0.2) min & 1523.0 & 10.9 $\pm$ 0.6 & 5.3 $\pm$ 0.1 \\
~~~~~~~~~~~& $^{120m}$I & (53 $\pm$ 4) min & 654.5 & 2.1 $\pm$ 0.7 & 10.6 $\pm$ 0.2  \\
\end{tabular}
\end{ruledtabular}
\end{table*}

\subsection{Irradiations}

Activation measurements of the \mbox{$^{120}$Te($p$,$\gamma$)$^{121}$I} and
\mbox{$^{120}$Te($p,n$)$^{120}$I} cross sections were performed at the
University of Notre Dame (Indiana, USA) Tandem Van de Graaff
accelerator. The accelerator provided a proton beam with energies ranging from
\mbox{2.5} to \mbox{8.0 MeV}, in steps of \mbox{0.5 MeV}, in the
laboratory frame. These energies correspond to effective center-of-mass
energies between \mbox{2.47} and \mbox{7.93 MeV}. The effective
center-of-mass energies ($E_{\rm c.m.}^{eff}$) are the proton
center-of-mass energies at which one half of the reaction yield
for the entire target thickness is obtained \cite{{Ili07},
{Rofls88}}.  This energy range covers the Gamow window for the
\mbox{$^{120}$Te($p$,$\gamma$)$^{121}$I} reaction at a temperature
of \mbox{3 x 10$^9$ K.} A schematic diagram of the target
irradiation chamber, located at the end of the beam line, is
illustrated in Fig. 1. The incident beam current was measured with
a Faraday cup mounted directly after the target chamber, and
isolated from the entire beam line. A secondary electron
suppression voltage of \mbox{-300 V} was applied at the entrance
of the Faraday cup. The beam was defined by an upstream collimator
with a diameter of \mbox{10 mm} and a smaller collimator with 5 mm
diameter at the target position mounted on a moveable target holder
with two positions; one for the collimator and the other for the target.
The beam was tuned by minimizing the current on both collimators and
using a quartz viewer at the end of the beam line.

\begin{figure}
\resizebox{0.35\textwidth}{!}{%
\includegraphics[180,120] [680,500]{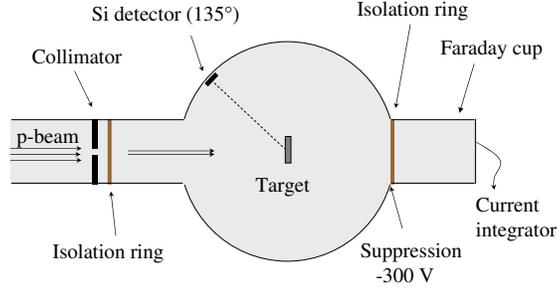}}
\caption{ A drawing of the components used in the beam line during
the irradiation. The beam was defined by a upstream collimator with a
diameter of \mbox{10 mm} and a smaller collimator with 5 mm diameter at target position.
The Si detector was placed at 135$^{\circ}$ with
respect to beam direction for RBS measurements. }
\end{figure}

The stability of the beam current was monitored with an integrator
in time intervals of \mbox{1 s}. Since the irradiation time period was
divided into segments that were sufficiently small, the beam intensity
was assumed constant over each segment. The calculation of the number of
reaction products in any time segment is discussed in
Refs. \cite{{Ozkan02}, {Famiano08}}. The applied current was between
\mbox{80} and \mbox{320 nA,} based on the thickness of the targets
and beam energy. The target was irradiated for \mbox{6 h} for the
lowest beam energy, \mbox{2.5 MeV.} With the increase of beam
energy, the irradiation time decreased, with a minimum time of
\mbox{15 min}, because of increasing cross section with energy.

After each irradiation, the target was removed from the target
chamber and then transported to the off-line gamma counting system
in order to measure the yield of the characteristic $\gamma$
activity of the produced unstable isotopes, \mbox{$^{121}$I and
$^{120}$I.}

\subsection{Determination of the activity}

The counting system was similar to that used previously to measure
the $\alpha$ captures of $^{112}$Sn \cite{Ozkan07} and $^{106}$Cd
\cite{Gyu06}. The $\gamma$ detection setup was composed of two
Clover detectors placed face-to-face in close geometry, \mbox{4.9 mm}
apart. In order to reduce the X-rays from the decays, a \mbox{0.59 mm}
thick Cu plate was placed in front of each Clover, so that more
than \mbox{99\%} of the \mbox{50 keV} X-rays were suppressed.
The whole assembly was shielded against room background with \mbox{5 cm}
of Pb and an inner \mbox{3 mm} Cu lining.

The irradiated target was placed in a plexiglass holder, and then
inserted into the fixed \mbox{4.9 mm} gap between the Clovers in
order to firmly constrain the center of the detection system for
reproducibility of the detection geometry. The $\gamma$ counting for
each run lasted between \mbox{1} and \mbox{7 h}, based on the
counting statistics. Each of the crystals were counted
individually (the so-called direct mode) in order to decrease
pileup and summing \cite{Saed}. The energies of the crystals were
recorded event by event together with the time of the event. A
pulser with a frequency \mbox{100 Hz} was fed into one of the Ge
preamplifiers, so that the dead time could be reconstructed
as a function of time. Dead time corrections were performed by
dividing the decay into sufficiently small time intervals depending
on the count rates.

In such a close geometry, which covers a solid angle of nearly \mbox{4$\pi$,}
the detection efficiency is relatively high, hence the correction for
coincidence summing effects becomes important. Summing correction
factors (between 2\% and 14\%) were determined by means of summing coefficients modified
from Ref. \cite{Shima}.

The photo-peak efficiency of the detection system was measured by
the efficiency-ratio method, which is described in Refs.
\cite{Ozkan07} and \cite{Debertin}. The relative efficiencies were
obtained with an uncalibrated $^{152}$Eu source (with respect to
the \mbox{245 keV} line $\gamma$ efficiency), and normalized to
the absolute photo-peak efficiency of a calibrated $^{137}$Cs
source. In previous works \cite{{Ozkan07}, {Gyu06}}, the
coincidence method was also used and found to be in agreement with
the efficiency-ratio method within the
uncertainties. The nearly \mbox{4$\pi$} detection geometry allowed
the angular correlation to be neglected. The absolute photo-peak
efficiencies of the $\gamma$ transitions used for the products of
the investigated reactions,
\mbox{$^{120}$Te($p$,$\gamma$)$^{121}$I} and
\mbox{$^{120}$Te($p,n$)$^{120}$I,} are given in Table I.

\begin{figure} [b]
\resizebox{0.53\textwidth}{!}{%
\includegraphics {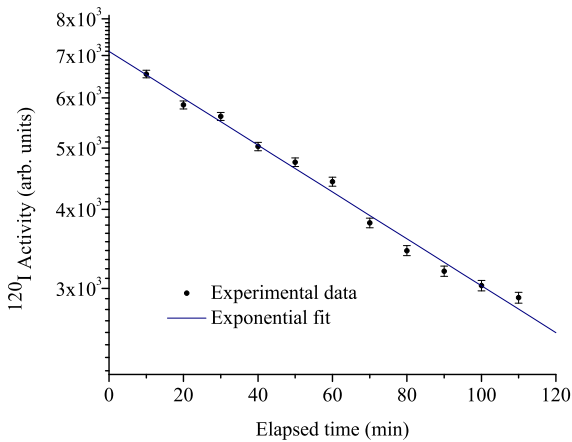}}
\caption{Decay of $^{120g}$I counted for 110 min with 10 min
intervals after \mbox{22 min} irradiation of $^{120}$Te at a proton
beam energy of \mbox{6.44 MeV.}
The solid line is the exponential fit to the measurement
($\chi^{2}$= 0.99).}
\end{figure}

In order to confirm that the analysis of the $\gamma$ transitions
of interest is realistic, the $\gamma$ peaks are confirmed not
only by their energy values but also by their known decay lifetimes.
The decays of the reaction products were plotted against time, and
their semi-logarithmic graphs are linear within the uncertainties.
This technique confirms that the analyzed $\gamma$ rays are
correctly identified and are associated with the isotopes of
interest. As an example, a decay of $^{120}$I counted for
\mbox{110 min} with \mbox{10 min} intervals following a \mbox{22 min}
irradiation of $^{120}$Te with \mbox{6.44 MeV} protons is shown
in \mbox{Fig. 2.} The $^{120g}$I decay curve through
the \mbox{1523 keV} line gives a half-life of \mbox{(81.3 $\pm$
1.5) min,} which is confirmed by the value taken from the
literature (Table I).

\section{Results and discussion}

The $^{120}$Te($p$,$\gamma$)$^{121}$I and $^{120}$Te($p,n$)$^{120}$I cross sections
have been measured in the effective center-of-mass energies
between \mbox{2.47} and \mbox{7.93 MeV} in order to test the consistency of
statistical model cross section predictions for $p$-process
nucleosynthesis simulations. The astrophysical $S$ factors have
also been determined from the measured cross sections. The
experimental energy range covers the Gamow window \mbox{(2.43} to
\mbox{4.64 MeV} at \mbox{3 x 10$^9$ K)}.
The results for the \mbox{$^{120}$Te($p$,$\gamma$)$^{121}$I} and
\mbox{$^{120}$Te($p,n$)$^{120}$I} reactions are summarized in
\mbox{Tables II} and III, respectively.

The uncertainty in the results stems from the following partial
errors: target thickness \mbox{($\sim$ 9\%)}, counting statistics
\mbox{(0.2\%} to \mbox{11\%)}, detection efficiency \mbox{(1.9\%}
to \mbox{2.4\%),} decay parameters \mbox{(0.2\%} to
\mbox{33\%)}, and beam current normalization (less than \mbox{2\%}).
The uncertainties in the effective center-of-mass
energies range between \mbox{0.02\%} and \mbox{0.5\%}; they were
calculated with the SRIM code \cite{SRIM} based on the proton
energy loss in the targets. To test for systematic uncertainties,
the \mbox{$^{120}$Te($p$,$\gamma$)$^{121}$I} reaction cross
section was measured at \mbox{3.5} and \mbox{5.0 MeV} using both
targets whose results are in excellent agreement (Table II).

\begin{table}[b]
\caption{Measured cross sections and $S$ factors of the
$^{120}$Te($p$,$\gamma)^{121}$I reaction.}
\begin{ruledtabular}
\setlength{\extrarowheight}{0.1cm}
\begin{tabular}{cr@{\hspace{0.15cm}}lr@{\hspace{0.15cm}$\pm$\hspace{-0.25cm}}
lr@{\hspace{0.15cm}$\pm$\hspace{-0.25cm}}l} $E$$_{\rm beam}$ &
\multicolumn{2}{c}{\hspace{-0.2cm}$E_{\rm c.m.}^{eff}$} &
\multicolumn{2}{c}{\hspace{-0.4cm}Cross section} &
 \multicolumn{2}{c}{\hspace{-0.5cm}$S$ factor} \\
{(MeV)} & \multicolumn{2}{c}{\hspace{-0.2cm}(MeV)} &
\multicolumn{2}{c}{\hspace{-0.4cm} (mb)} &
\multicolumn{2}{c}{\hspace{-0.5cm}(10$^{11}$ keV b)}\\
\hline
2.500 & 2.467$\pm$0.013 &  & 0.0023 & 0.0003& 7.62 & 1.08\\
3.000 & 2.963$\pm$0.012 & & 0.030 & 0.003 & 7.15 & 0.73 \\
3.500 & 3.460$\pm$0.011 &  & 0.194 & 0.019 & 5.91 & 0.58 \\
3.500 & 3.468$\pm$0.003 &  & 0.200 & 0.020 & 6.01 & 0.51 \\
4.000 & 3.958$\pm$0.010 &  & 0.706 & 0.069 & 4.16 &  0.41 \\
4.500 & 4.454$\pm$0.009 &  & 2.55 & 0.25 & 3.89 & 0.38 \\
5.000 & 4.950$\pm$0.008 & & 5.68 & 0.56 & 2.77 & 0.27\\
5.000 & 4.956$\pm$0.003 & & 5.63 & 0.55 & 2.75 & 0.22\\
5.500 & 5.452$\pm$0.002 && 18.7 & 1.8 & 3.43 & 0.34 \\
6.000 & 5.942$\pm$0.008 &  & 34.3 & 3.3 & 2.71 & 0.27 \\
6.500 & 6.444$\pm$0.002 & & 46.5 & 4.5 & 1.74 & 0.17 \\
7.000 & 6.940$\pm$0.002 & &  23.2 & 2.3 & 0.45 & 0.04 \\
7.500 & 7.436$\pm$0.002& & 21.8 & 2.2 & 0.24 & 0.02 \\
8.000 & 7.932$\pm$0.002 & & 16.5 & 1.7 & 0.10 & 0.01 \\
\end{tabular}
\end{ruledtabular}
\end{table}

\begin{table}
\caption{Measured cross sections and $S$ factors of the
$^{120}$Te($p,n$)$^{120}$I reaction.}
\begin{ruledtabular}
\setlength{\extrarowheight}{0.1cm}
\begin{tabular}{cr@{\hspace{0.15cm}}lr@{\hspace{0.15cm}$\pm$\hspace{-0.25cm}}
lr@{\hspace{0.15cm}$\pm$\hspace{-0.25cm}}l} $E$$_{\rm beam}$ &
\multicolumn{2}{c}{\hspace{-0.2cm}$E_{\rm c.m.}^{eff}$} &
\multicolumn{2}{c}{\hspace{-0.4cm}Cross section} &
 \multicolumn{2}{c}{\hspace{-0.5cm}$S$ factor} \\
{(MeV)} & \multicolumn{2}{c}{\hspace{-0.2cm}(MeV)} &
\multicolumn{2}{c}{\hspace{-0.4cm}(mb)} &
\multicolumn{2}{c}{\hspace{-0.5cm}(10$^{11}$ keV b)}
\\ \hline 6.500 & 6.444$\pm$0.002 & & 20.6 & 2.1& 0.77 & 0.08
\\ 7.000 & 6.940$\pm$0.002 & & 72.6 & 7.5 & 1.41 & 0.14
\\ 7.500 & 7.436$\pm$0.002 & & 133 & 13 & 1.44 & 0.14
\\ 8.000 & 7.932$\pm$0.002 & & 178 & 18 & 1.12 & 0.11
\\
\end{tabular}
\end{ruledtabular}
\end{table}

\begin{table}
\caption{Measured cross sections of the $^{120}$Te($p,n$)$^{120}$I reactions
that produce ground $^{120g}$I and isomeric $^{120m}$I states. For
the analysis, 1523 keV and 654.5 keV $\gamma$ transitions,
respectively, were used: for the decay parameters see Table I.}
\begin{ruledtabular}
\setlength{\extrarowheight}{0.1cm}
\begin{tabular}{lccc}
\multicolumn{1}{c}{$E$$_{\rm beam}$} &
\multicolumn{1}{c}{$E_{\rm c.m.}^{eff}$} &
\multicolumn{2}{c}{Cross Section (mb)} \\
\multicolumn{1}{c} {(MeV)} &
\multicolumn{1}{c} {(MeV)} &
\multicolumn{1}{c}{$^{120g}$I (1523 keV) } &
\multicolumn{1}{c}{$^{120m}$I (654.5 keV)}  \\
\hline
6.500 & 6.444$\pm$0.002 &  19.1$\pm$2.0 & 1.53$\pm$0.66
\\7.000 & 6.940$\pm$0.002 &  65.4$\pm$6.8 & ~7.2$\pm$3.0
\\7.500 & 7.436$\pm$0.002 &  123$\pm$13 & 10.8$\pm$4.5
\\8.000 & 7.932$\pm$0.002 &  160$\pm$16 & 18.1$\pm$7.5
\\
\end{tabular}
\end{ruledtabular}
\end{table}

\begin{table*}
\caption{Model calculations using TALYS and NON-SMOKER codes with various combinations of the
nuclear parameters: nuclear level densities (NLD), and optical model parameters (OMP).
JLM model used in TALYS and NON-SMOKER is with the modification of \cite{Bauge01} and \cite{Lej80}, respectively.}
\setlength{\extrarowheight}{0.1cm}
\centering
\begin{ruledtabular}
\begin{tabular}{lccc}
\parbox[t]{1.5cm}{\centering{Label}} &
\parbox[t]{2.0cm}{\centering{NLD-model}} &
\parbox[t]{1.5cm}{\centering{OMP}} &
 \\
\hline
TALYS-BSFG-JLM & BSFG \cite{Talys01}& JLM \cite{{JLM77}, {Bauge01}} \\
TALYS-CTFG-JLM & CTFG \cite{Talys01} & JLM \cite{{JLM77}, {Bauge01}} \\
TALYS-BSFG-KD & BSFG \cite{Talys01} & KD \cite{Koning03} \\
TALYS-default & CTFG \cite{Talys01} & KD \cite{Koning03} \\
NON-SMOKER-standard & BSFG \cite{Raus01} & JLM \cite{{JLM77}, {Lej80}} \\
\end{tabular}
\end{ruledtabular}
\end{table*}

The $^{120}$Te($p,n$)$^{120}$I reaction produces ground
($^{120g}$I) and isomeric ($^{120m}$I) states of $^{120}$I with
the half-lives \mbox{81.6 min} and \mbox{53 min,} respectively.
The cross section of this reaction was determined by summing the
partial cross sections of the \mbox{$^{120}$Te($p,n$)$^{120g}$I}
and \mbox{$^{120}$Te($p,n$)$^{120m}$I} reactions. The partial
cross sections of these \mbox{($p,n$)} reactions were measured by
using the individual decay parameters (Table I) of the ground and
isomeric states. For the analysis of \mbox{($p,n$)} data, the
\mbox{1523 keV} ($^{120g}$I) and \mbox{654.5 keV} ($^{120m}$I)
$\gamma$ transitions were chosen because these two $\gamma$ transitions
are associated exclusively with the decay of ground and isomeric states.
The cross sections of these two \mbox{($p,n$)}
reactions are listed separately in the \mbox{Table IV.}

The measured cross sections and astrophysical $S$ factors of the $^{120}$Te($p$,$\gamma$)$^{121}$I
and $^{120}$Te($p,n$)$^{120}$I reactions
have been compared with the Hauser-Feshbach statistical model
calculations obtained with the standard settings of the statistical
model code NON-SMOKER \cite{{Raus00}, {Raus01}} and TALYS \cite{Talys01} with
the default parameters \cite{{Koning03}, {Ericson60}, {Gilbert65}} (and also with parameters similar to those of the standard
settings of the NON-SMOKER \cite{{Raus97}, {JLM77}, {Lej80}},
as discussed later). The default optical model
potentials used in TALYS are the local and global parameterizations
for protons from Ref. \cite{Koning03}. For the nuclear level
density, the TALYS code uses the parametrization by Refs.
\cite{Ericson60} and \cite{Gilbert65}.
For the standard settings of the statistical code NON-SMOKER, the optical potential
is the widely-used semimicroscopic potential of Jeukenne $et$ $al$. \cite{JLM77}
(JLM) with the low energy modifications by \cite{Lej80} and the
nuclear level densities of Rauscher $et$ $al$. \cite{Raus97}.

Figure 3 shows that good agreement is generally observed for the
\mbox{$^{120}$Te($p$,$\gamma$)$^{121}$I} cross section values,
although it seems the theoretical calculations using the standard
NON-SMOKER and TALYS codes deviate considerably from the data at
energies above the neutron threshold. The fact that the NON-SMOKER
code underestimates the ($p,n$) measurements may explain why
NON-SMOKER predictions overestimate the ($p$,$\gamma$) data at
energies above \mbox{6.44 MeV}. The predicted $S$ factor
values overestimate the data by approximately a factor of 2.5
above \mbox{6.44 MeV} while they deviate from the experimental
values by factors of less than 1.7 at energies below \mbox{6.44 MeV},
for both standard codes, as seen in \mbox{Fig. 4}. Three points in
the \mbox{($p$,$\gamma$)} data at energies between \mbox{3.96} and
\mbox{4.95 MeV} are lower by a factor of 1.3 to 1.7 compared to
the predictions (Fig. 4). That is well within the uncertainty
range defined by the standard input parameters for the Hauser-Feshbach models.
Similar behavior is seen in some
\mbox{($p$,$\gamma$)} measurements whose references are listed in
the \mbox{Sec. I}, especially for the
\mbox{$^{86}$Sr($p$,$\gamma$)$^{87}$Y} reaction \cite{Gyu01}.

\begin{figure} [b]
\resizebox{0.53\textwidth}{!}{%
\includegraphics {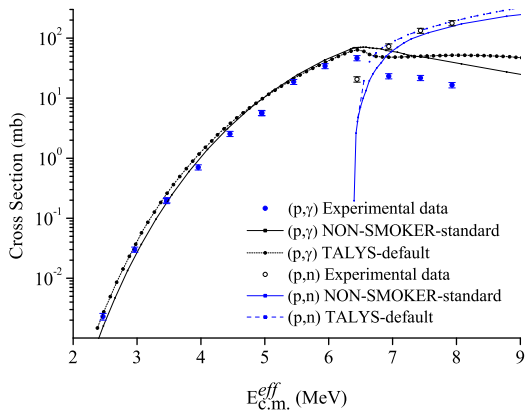}}
\caption{ (Color online) The experimental cross section results of
$^{120}$Te($p$,$\gamma)^{121}$I and $^{120}$Te($p,n$)$^{120}$I
reactions in comparison with standard
NON-SMOKER \cite{Raus01} and TALYS predictions with the default parameters \cite{Talys01}.}
\end{figure}

For the $^{120}$Te($p$,$\gamma$)$^{121}$I reaction, the energy
dependence of the astrophysical $S$ factor is slightly better
described by the statistical model code TALYS with the default
parameters at energies below \mbox{6.94 MeV} (Fig. 4). The
difference between the predictions of the two model codes can
mainly be attributed to the different proton optical potential and
nuclear level densities used in the codes. In order to try and
identify the source of the observed discrepancies between the two
calculations, the optical model potentials (OMP) and nuclear level
densities (NLD) of TALYS were varied. Various combinations of the
nuclear parameters, and their labels, used in the model codes are
given in \mbox{Table V}.

The impact of different parameters (as indicated with
\mbox{TALYS-BSFG-JLM} for the TALYS $S$ factor predictions) is shown
in \mbox{Fig. 4} compared to the predictions of NON-SMOKER for
its standard settings.  The experimental
\mbox{$^{120}$Te($p$,$\gamma$)$^{121}$I} $S$ factor results are
within the uncertainty range resulting from the choice of
different parameter settings. \mbox{Figure 4} also shows that the
predictions of both TALYS and NON-SMOKER using similar parameters
(standard NON-SMOKER) have the same energy dependence. However,
\mbox{TALYS-BSFG-JLM} is generally lower than the standard
NON-SMOKER prediction by a factor of 2. In order to
investigate which parameters cause the changes in magnitude, only
one parameter of the default TALYS was changed each time. The
changes in the OMP and NLD are labeled TALYS-CTFG-JLM and
TALYS-BSFG-KD, respectively, in \mbox{Fig. 4.} The results indicate
that the OMP is the most critical parameter for the setting of
magnitude. Indeed, TALYS uses the optical nucleon potential of
Jeukenne $et$ $al$. \cite{JLM77} (with the modification of Bauge
$et$ $al$. \cite{Bauge01}) without the low-energy modifications,
unlike the standard NON-SMOKER (low energy modifications by
\cite{Lej80}). It should be also emphasized that both codes apply
the constant temperature formula \cite{Gilbert65} for the nuclear
level density in order to correct the behavior due to the
divergence of the Fermi-gas model at very low excitation energies.
As the energy increases, the results of TALYS-BSFG-KD predictions
deviate from the ones using TALYS with its standard default
parameters and shows a similar energy dependence as compared to the standard
NON-SMOKER predictions.

\begin{figure} [t]
\resizebox{0.53\textwidth}{!}{%
\includegraphics {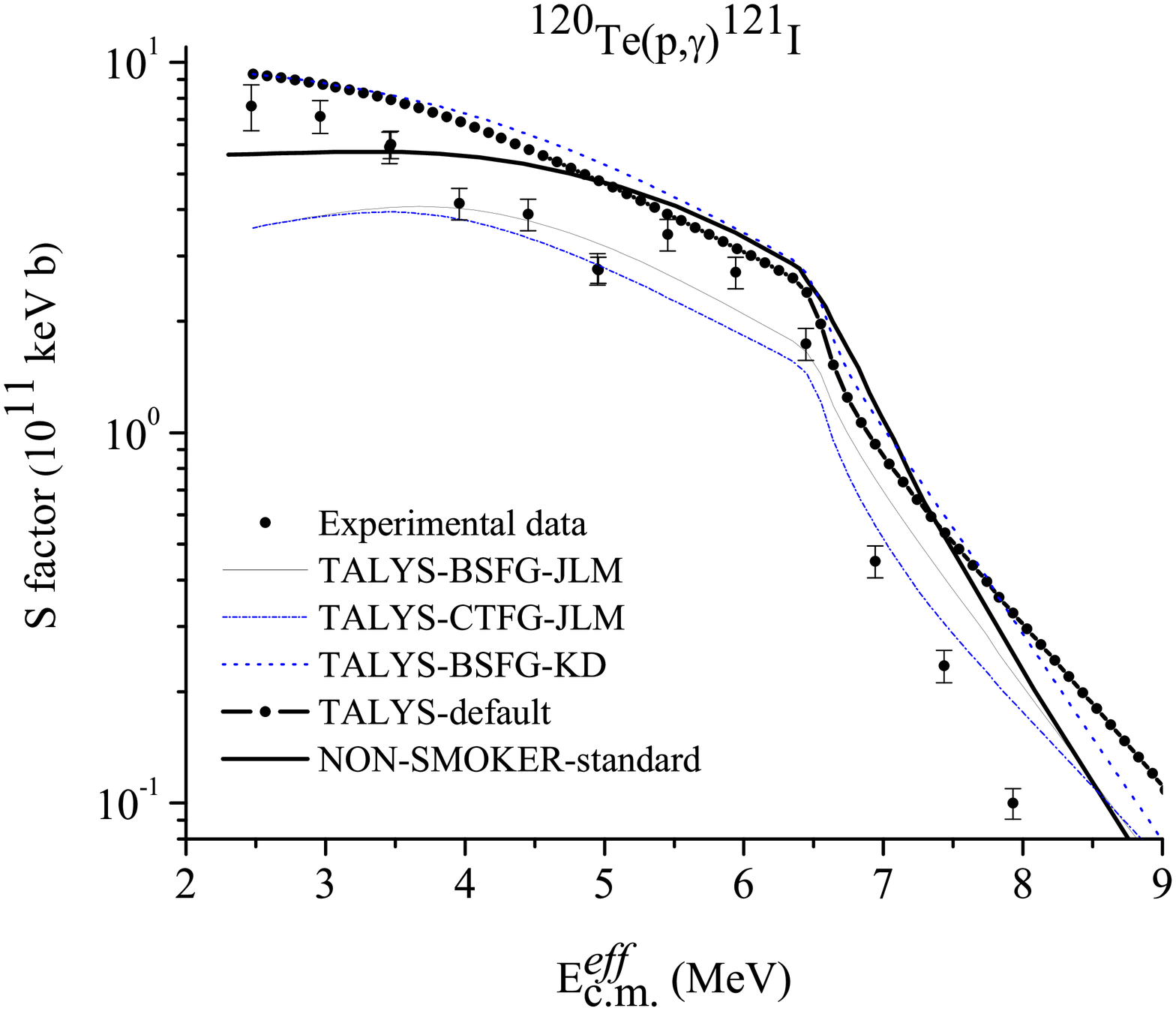}}
\caption{ (Color online) Comparison of predicted astrophysical
$S$ factors of $^{120}$Te($p$,$\gamma)^{121}$I reaction with four different
TALYS code calculations, and standard NON-SMOKER code,
using the combinations of nuclear parameters described in \mbox{Table V.}
The experimental $S$ factors of the \mbox{($p$,$\gamma$)} reaction are also presented. The
JLM model used in TALYS and NON-SMOKER is with the modification of \cite{Bauge01} and \cite{Lej80}, respectively.}
\end{figure}

\begin{figure} [t]
\resizebox{0.53\textwidth}{!}{%
\includegraphics {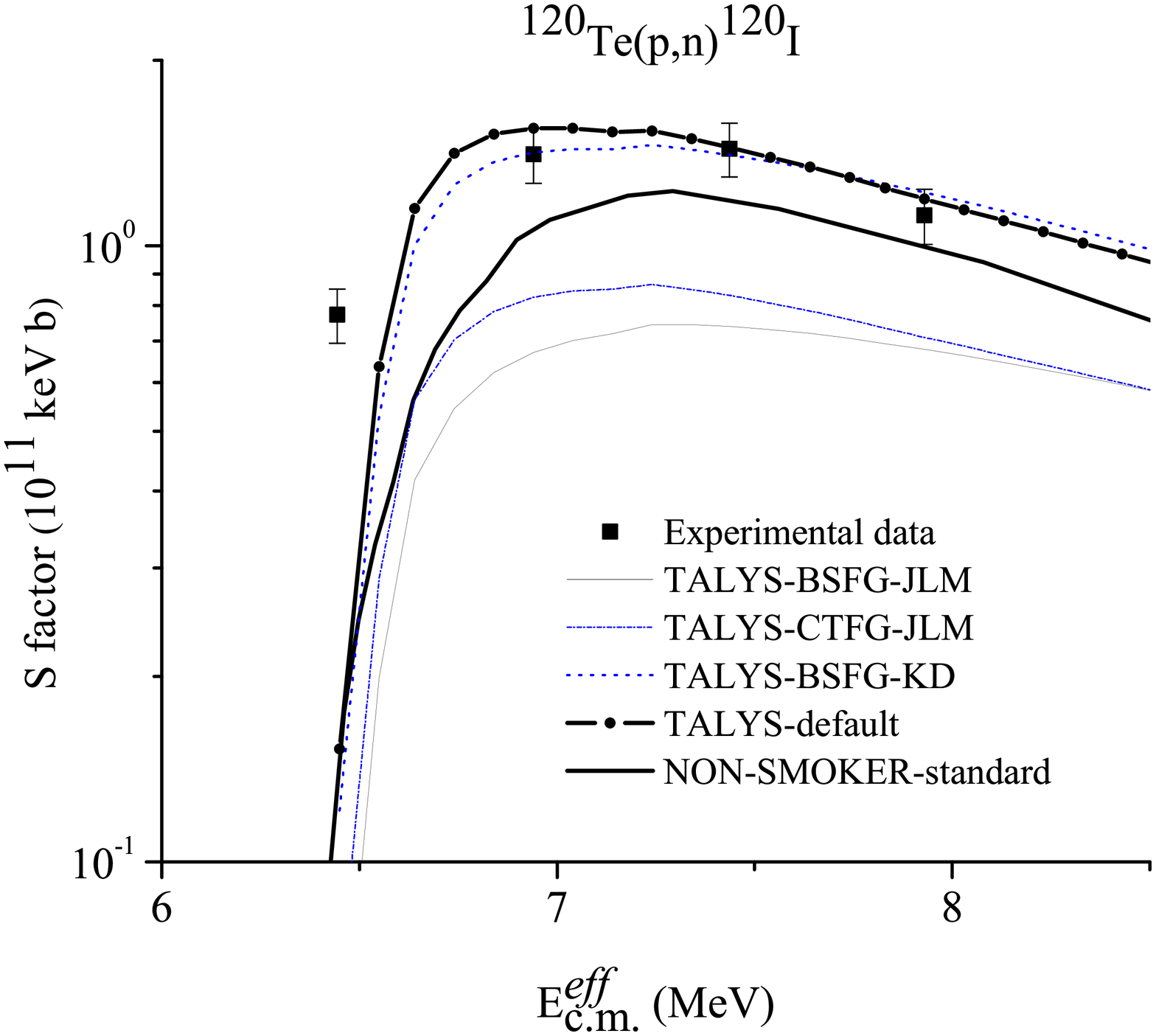}}
\caption{ (Color online) Comparison of predicted astrophysical
$S$ factors of $^{120}$Te($p,n$)$^{120}$I reaction with four different
TALYS code calculations, and standard NON-SMOKER code,
using the combinations of nuclear parameters described in \mbox{Table V.}
The experimental $S$ factors of the \mbox{($p,n$)} reaction are also presented. The
JLM model used in TALYS and NON-SMOKER is with the modification of \cite{Bauge01} and \cite{Lej80}, respectively.}
\end{figure}

In the case of the $^{120}$Te($p,n$)$^{120}$I reaction, a better
agreement is found between the experimental data and both standard
codes (Figs. 3 and 5). The NON-SMOKER calculations of the
astrophysical $S$ factors underestimate the measured values by
factors less than 1.4  while the TALYS predictions and the data
for the studied energies are in excellent agreement, with the
exception of the point at \mbox{6.44 MeV}, as seen in \mbox{Fig.
5.} The predicted $S$ factor of NON-SMOKER and TALYS at this
particular energy are below the experimental value by a factor of 5.
The \mbox{($p,n$)} channel opens around this energy
($Q$-value of \mbox{6.397 MeV} corresponding to the threshold value
of \mbox{6.451 MeV).} The fact that the codes use the
experimentally known $Q$-value with \mbox{20 keV} uncertainty
could explain, in part, the difference at the lowest measured energy
for the \mbox{($p,n$)} reaction because the cross section decreases
steeply at energies near the threshold value.

The same variations in the nuclear parameters \mbox{(Table V)} of
the code TALYS were applied to the ($p,n$) reaction. \mbox{Figure
5} shows that the $S$ factor results of NON-SMOKER provide better
agreement with the experimental data than those of TALYS when the
standard NON-SMOKER parameters are used (as indicated with
\mbox{TALYS-BSFG-JLM}). The model calculations are more sensitive
to the choice of OMP than the other nuclear parameters, as seen in
\mbox{Fig. 5}.

\section{Summary and conclusions}

The total cross sections of the reactions
\mbox{$^{120}$Te($p$,$\gamma$)$^{121}$I} and \mbox{$^{120}$Te($p,n$)$^{120}$I}
have been measured via the activation method, and their astrophysical
$S$ factors have been derived in the effective center-of-mass
energy range between \mbox{2.47} and \mbox{7.93 MeV.}

Measurements of the cross sections were performed for a broad
energy range, covering the Gamow window centered at \mbox{3.53
MeV} for \mbox{T = 3 x 10$^9$ K}. It has been pointed out that in
cases where a nuclear reaction proceeds through narrow resonances,
an effective stellar energy window can differ significantly from
the commonly used Gamow peak \cite{Newton07}. However, for heavier
nuclei, the nuclear level density is high, and the cross section
is characterized by a multitude of overlapping resonances which is
eligible for the statistical Hauser-Feshbach treatment
\cite{Raus97}.

The experimental results of astrophysical $S$ factors for the
\mbox{$^{120}$Te($p$,$\gamma$)$^{121}$I} and
\mbox{$^{120}$Te($p,n$)$^{120}$I} reactions have been compared
with the predictions of statistical model calculations using the
standard NON-SMOKER code as well as the TALYS code with various
combinations of the nuclear parameters (listed in Table V). The
discrepancies in the results between the predictions are
relatively small and can be attributed to the choice of nuclear
input parameters used in the codes. The $S$ factor results are
more sensitive to the OMP than to the NLD in the astrophysically
relevant low-energy region, for both ($p$,$\gamma$) and ($p,n$)
reactions. The best overall agreement is obtained with the OMPs of
KD \cite{Koning03} using the code TALYS.

For the $^{120}$Te($p$,$\gamma$)$^{121}$I reaction, the default setting of
TALYS (-CTFG-KD) offers a better reproduction below the ($p,n$) threshold.
Changing the OMP of TALYS to JLM \cite{JLM77} gives a poor reproduction as a whole.
The deviations at the very lowest energies are expected because OMP is parameterized
as a function of energy over a broader mass region \cite{Gyu03}. It should be also
emphasized that the ($p$,$\gamma$) $S$ factors are very sensitive to the proton width at
the lowest energies \cite{Kiss07,{Gyu07}}.

In the case of the \mbox{$^{120}$Te($p,n$)$^{120}$I} reaction, the
code TALYS with the default parameters is able to reproduce the
data very well except for the point at 6.44 MeV. This has been
identified as a threshold effect.
The choice of OMPs clearly plays a critical role over the entire
energy region. It should be emphasized that ($p,n$) reactions are
also sensitive to the proton width at all energies
\cite{Kiss07,{Gyu07}}.

Overall, the results of the experiments indicate good agreement
with the theoretical predictions within the uncertainty range of
the nuclear structure input parameters for the model predictions.
This confirms earlier observations of $p$-process reaction
measurements on lower Z targets and confirms the validity of
Hauser-Feshbach model applications for $p$-process reactions over
the entire mass range from Z=35 to Z=52. The results for the
specific proton capture reaction on $^{120}$Te suggests that the
Hauser-Feshbach predictions for the inverse stellar photo
disintegration process \cite{Raus04} are also reliable within the
model uncertainty range.

\begin{acknowledgments}
This work was supported by The Scientific and Technological Research
Council of Turkey TUBITAK-Grant-108T508,
Kocaeli University BAP- Grant-2007/36,
the National Science Foundation NSF-Grant-0434844,
the Joint Institute for Nuclear Astrophysics JINA (www.JINAweb.org) PHY02-16783, and The Hungarian Scientific Research Fund Programs OTKA (K68801, T49245).
\end{acknowledgments}

\bibliography{apssamp}

\end{document}